\documentclass{amsart}
\usepackage{amssymb}
\usepackage{amscd}

\begin{document}
\title{QUANTUM ENTANGLEMENTS AND ENTANGLED MUTUAL\ ENTROPY}
\author{Viacheslav P Belavkin}
\address{Department of Mathematics, University of Nottingham, NG7 2RD Nottingham, UK}
\thanks{The first author acknowledge the hospitality of Tokyo University of Science
during his stay at the JSPS Senior Fellowship in 1998.}
\author{Masanori Ohya}
\address{Department of Information Sciences, Science University of Tokyo, 278 Noda
City, Chiba, Japan}
\date{July 20, 1998}
\subjclass{Quantum Probability and Information}
\thanks{The first author is grateful for the support under the JSPS Invitation
Fellowship Program for Research in Japan.}
\maketitle

\begin{abstract}
The mathematical structure of quantum entanglement is studied and classified
from the point of view of quantum compound states. We show that the
classical-quantum correspondences such as encodings can be treated as
diagonal (d-) entanglements. The mutual entropy of the d-compound and
entangled states lead to two different types of entropies for a given
quantum state: the von Neumann entropy, which is achieved as the supremum of
the information over all d-entanglements, and the dimensional entropy, which
is achieved at the standard entanglement, the true quantum entanglement,
coinciding with a d-entanglement only in the case of pure marginal states.
The q-capacity of a quantum noiseless channel, defined as the supremum over
all entanglements, is given by the logarithm of the dimensionality of the
input algebra. It doubles the classical capacity, achieved as the supremum
over all d-entanglements (encodings), which is bounded by the logarithm of
the dimensionality of a maximal Abelian subalgebra.
\end{abstract}

\section{Introduction}

Recently, the specifically quantum correlations, called in quantum physics
entanglements, are used to study quantum information processes, in
particular, quantum computation, quantum teleportation, quantum cryptography 
\cite{Ben93,Eke,JS}. There have been mathematical studeis of the
entanglements in \cite{BBPSSW,Maj,Sch1}, in which the entangled state is
defined by a state not written as a form $\sum_{k}\lambda _{k}\rho
_{k}\otimes \sigma _{k}$ with any states $\rho _{k}$ and $\sigma _{k}.$
However it is obvious that there exist several correlated states written as
separable forms above. Such correlated, or entangled states have been also
discussed in several contexts in quantum probability such as quantum
measurement and filtering \cite{Bel80, Bel94}, quantum compound state\cite
{Ohy83,Ohy83-2} and lifting \cite{ AcO}. In this paper, we study the
mathematical structure of quantum entangled states to provide a finer
classification of quantum sates, and we discuss the informational degree of
entanglement and entangled quantum mutual entropy.

We show that the entangled states can be treated as generalized compound
states, the nonseparable states of quantum compound systems which are not
represetable by convex combinations of the product states. The compound
states, called o-entangled, are defined by orthogonal decompositions of
their marginal states. This is a particular case of so called separable
state of a compound system, the convex combination of the product states
which we call c-entangled. The o-entangled compound states are most
informative among c-entangled states in the sense that the maximum of mutual
entropy over all c-entanglements to the quantum system $\mathcal{A}$ is
achieved on the extreme o-entangled states as the von Neumann entropy $%
S\left( \varrho \right) $ of a given normal state $\varrho $ on $\mathcal{A}$
. Thus the maximum of mutual entropy over all classical couplings, described
by c-entanglements of (quantum) probe systems $\mathcal{B}$ to the system $%
\mathcal{A}$, is bounded by $\ln \mathrm{rank}\mathcal{A}$, the logarithm of
the rank of the von Neumann algebra $\mathcal{A}$, defined as the
dimensionality of the maximal Abelian subalgebra $\mathcal{A}^{\circ
}\subseteq \mathcal{A}$. Due to $\dim \mathcal{A}\leq \left( \mathrm{rank}%
\mathcal{A}\right) ^{2}$, it is achieved on the normal tracial $\rho =\left( 
\mathrm{rank}\mathcal{A}\right) ^{-1}I$ only in the case of finite
dimensional $\mathcal{A}$.

More general than o-entangled states, the d-entangled states, are defined as
c-entangled states by orthogonal decomposition of only one marginal state on
the probe algebra $\mathcal{B}$. They can give bigger mutual entropy for a
quantum noisy channel than the o-entangled state which gains the same
information as d-entangled extreme states in the case of a deterministic
channel.

We prove that the truly (strongest) entangled states are most informative in
the sense that the maximum of mutual entropy over all entanglements to the
quantum system $\mathcal{A}$ is achieved on the quasi-compound state, given
by an extreme entanglement of the probe system $\mathcal{B}=\mathcal{A}$
with coinciding marginals, called standard for a given $\varrho $. The
standard entangled state is o-entangled only in the case of Abelian $%
\mathcal{A}$ or pure marginal state $\varrho $. The gained information for
such extreme q-compound state defines another type of entropy, the
quasi-entropy $\widetilde{S}\left( \varrho \right) $ which is bigger than
the von Neumann entropy $S\left( \varrho \right) $ in the case of
non-Abelian $\mathcal{A}$ (and mixed $\varrho $.) The maximum of mutual
entropy over all quantum couplings, described by true quantum entanglements
of probe systems $\mathcal{B}$ to the system $\mathcal{A}$ is bounded by $%
\ln \mathrm{\dim }\mathcal{A}$, the logarithm of the dimensionality of the
von Neumann algebra $\mathcal{A}$, which is achieved on a normal tracial $%
\rho $ in the case of finite dimensional $\mathcal{A}$. Thus the q-entropy $%
\widetilde{S}\left( \varrho \right) $, which can be called the dimensional
entropy, is the true quantum entropy, in contrast to the von Neumann entropy 
$S\left( \varrho \right) $, which is semi-classical entropy as it can be
achieved as a supremum over all couplings with the classical probe systems $%
\mathcal{B}$. These entropies coincide in the classical case of Abelian $%
\mathcal{A}$ when $\mathrm{rank}\mathcal{A}=\dim \mathcal{A}$. In the case
of non-Abelian finite-dimensional $\mathcal{A}$ the q-capacity $C_{q}=\ln 
\mathrm{\dim }\mathcal{A}$ is achieved as the supremum of mutual entropy
over all q-encodings (correspondences), described by entanglements. It is
strictly bigger then the classical capacity $C=\mathrm{\ln rank}\mathcal{A}$
of the identity channel, which is achieved as the supremum over usual
encodings, described by the classical-quantum correspondences $\mathcal{A}
^{\circ }\rightarrow \mathcal{A}$.

In this short paper we consider the case of a simple algebra $\mathcal{A}=%
\mathcal{L}\left( \mathcal{H}\right) $ for which some results are rather
obvious and given without proofs. The proofs are given in the complete paper 
\cite{BeO98} for a more general case of decomposable algebra $\mathcal{A}$
to include the classical discrete systems as a particular quantum case, and
will be published elsewhere.

\section{Compound States and Entanglements}

Let $\mathcal{H}$ denote the (separable) Hilbert space of a quantum system,
and $\mathcal{A}=\mathcal{L}\left( \mathcal{H}\right) $ be the algebra of
all linear bounded operators {on }$\mathcal{H}$. A bounded linear functional 
$\varrho :\mathcal{{\ A}\rightarrow }\mathbf{C}$ is called a state on $%
\mathcal{A}$ if it is positive (i.e., $\varrho \left( A\right) \geq 0$ for
any positive operator $A$ in $\mathcal{A}$) and normalized $\varrho (I)=1$
for the identity operator $I$ in $\mathcal{A}$ . A normal state can be
expressed as 
\begin{equation}
\varrho \left( A\right) =\mathrm{tr}_{\mathcal{G}}\kappa ^{\dagger }A\kappa =%
\mathrm{tr}A\rho ,\text{ \quad }A\in \mathcal{A}\text{.}  \label{1.1}
\end{equation}
In (2.1), $\mathcal{G}$ is another separable Hilbert space, $\kappa $ is a
linear Hilbert-Schmidt operator from $\mathcal{G}$ to $\mathcal{H}$ and $%
\kappa ^{\dagger }$ is the adjoint operator of $\kappa $ from $\mathcal{H}$
to $\mathcal{G}$. This $\kappa $ is called the amplitude operator, and it is
called just the amplitude if $\mathcal{G}$ is one dimensional space $\mathbb{%
C}$ , corresponding to the pure state $\varrho \left( A\right) =\kappa
^{\dagger }A\kappa $ for a $\kappa \in \mathcal{H}$ with $\kappa ^{\dagger
}\kappa =\Vert \kappa \Vert ^{2}=1,$ in which case $\kappa ^{\dagger }$ is
the adjoint functional from $\mathcal{H}$ to $\mathbb{C}$. Moreover the
density operator $\rho $ in (2.1) is $\kappa \kappa ^{\dagger }$ uniquely
defined as a positive trace class operator $\mathrm{P}_{\mathcal{A}}\in 
\mathcal{A}$ . Thus the predual space $\mathcal{A}_{*}$ can be identified
with the Banach space $\mathcal{T}\left( \mathcal{H}\right) $ of all trace
class operators in $\mathcal{H}$ (the density operators $\mathrm{P}_{%
\mathcal{A}}\in \mathcal{A}_{*}$, $\mathrm{P}_{\mathcal{B}}\in \mathcal{B}%
_{*}$ of the states $\varrho $, $\varsigma $ on different algebras $\mathcal{%
A}$, $\mathcal{B}$ will be usually denoted by different letters $\rho
,\sigma $ corresponding to their Greek variations $\varrho $, $\varsigma $.)

In general, $\mathcal{G}$ is not one dimensional, the dimensionality $\dim 
\mathcal{G}$ must be not less than $\mathrm{rank}\rho $, the dimensionality
of the range $\mathrm{ran}\rho \subseteq \mathcal{H}$ of the density
operator $\rho .$ We shall equip it with an isometric involution $%
J=J^{\dagger }$, $J^2=I$, having the properties of complex conjugation on $%
\mathcal{G}$, 
\begin{equation*}
J\sum \lambda _j\zeta _j=\sum \bar{\lambda _j}J\zeta _j,\quad \forall
\lambda _j\in \mathbf{C},\zeta _j\in \mathcal{G}
\end{equation*}
with respect to which $J\sigma =\sigma J$ for the positive and so
self-adjoint operator $\sigma =\kappa ^{\dagger }\kappa =\sigma ^{\dagger }$
on $\mathcal{G}$. The latter can also be expressed as the symmetricity
property $\tilde{\varsigma}=\varsigma $ of the state $\varsigma \left(
B\right) =$ $\mathrm{tr}B\sigma $ given by the real and so symmetric density
operator $\bar{\sigma}=\sigma =\tilde{\sigma}$ on $\mathcal{G}$ with respect
to the complex conjugation $\bar{B}=JBJ$ and the tilda operation ($\mathcal{G%
}$-transponation) $\tilde{B}=JB^{\dagger }J$ on the algebra $\mathcal{B}=%
\mathcal{L}\left( \mathcal{G}\right) $.

For example, $\mathcal{G}$ can be realized as a subspace of $l^2(\mathbf{N})$
of complex sequences $\mathbf{N}\ni n\mapsto \zeta \left( n\right) \in %
\mathbb{C}$, with $\sum_n\left| \zeta \left( n\right) \right| ^2<+\infty $
in the diagonal representation $\sigma =\left[ \mu \left( n\right) \delta
_n^m\right] $. The involution $J$ can be identified with the complex
conjugation $C\zeta \left( n\right) =\bar{\zeta}\left( n\right) $, i.e., 
\begin{equation*}
C:\zeta =\sum_n|n\rangle \zeta \left( n\right) \mapsto C\zeta
=\sum_n|n\rangle \bar{\zeta}\left( n\right)
\end{equation*}
in the standard basis $\left\{ |n\rangle \right\} \subset \mathcal{G}$ of $%
l^2(\mathbf{N})$. In this case $\kappa =\sum \kappa _n\langle n|$ is given
by orthogonal eigen-amplitudes $\kappa _n\in \mathcal{H}$, $\kappa
_m^{\dagger }\kappa _n=0$, $m\neq n$, normalized to the eigen-values $%
\lambda \left( n\right) =\kappa _n^{\dagger }\kappa _n=\mu \left( n\right) $
of the density operator $\rho $ such that $\rho =\sum \kappa _n\kappa
_n^{\dagger }$ is a Schatten decomposition, i.e. the spectral decomposition
of $\rho $ into one-dimensional orthogonal projectors. In any other basis
the operator $J$ is defined then by $J=U^{\dagger }CU$, where $U$ is the
corresponding unitary transformation. One can also identify $\mathcal{G}$
with $\mathcal{H}$ by $U\kappa _n=\lambda \left( n\right) ^{1/2}|n\rangle $
such that the operator $\rho $ is real and symmetric, $J\rho J=\rho =J\rho
^{\dagger }J$ in $\mathcal{G}=\mathcal{H}$ with respect to the involution $J$
defined in $\mathcal{H}$ by $J\kappa _n=\kappa _n$. Here $U$ is an isometric
operator $\mathcal{H}\rightarrow l^2\left( \mathbb{N}\right) $ diagonalizing
the operator $\rho $: $U\rho U^{\dagger }=\sum |n\rangle \lambda \left(
n\right) \langle n|$. The amplitude operator $\kappa =\rho ^{1/2}$
corresponding to $\mathcal{B}=\mathcal{A}$, $\sigma =\rho $ is called
standard.

Given the amplitude operator $\kappa $, one can define not only the states $%
\varrho $ $\left( \rho =\kappa \kappa ^{\dagger }\right) $and $\varsigma $ $%
\left( \sigma =\kappa ^{\dagger }\kappa \right) $on the algebras $\mathcal{A}
=\mathcal{L}\left( \mathcal{H}\right) $ and $\mathcal{B}=\mathcal{L}\left( 
\mathcal{G}\right) $ but also a pure entanglement state $\varpi $ on the
algebra $\mathcal{B}\otimes \mathcal{A}$ of all bounded operators on the
tensor product Hilbert space $\mathcal{G}\otimes \mathcal{H}$ by

\begin{equation*}
\varpi \left( B\otimes A\right) =\mathrm{tr}_{\mathcal{G}}\tilde{B}\kappa
^{\dagger }A\kappa =\mathrm{tr}_{\mathcal{H}}A\kappa \tilde{B}\kappa
^{\dagger }.
\end{equation*}
Indeed, thus defined $\varpi $ is uniquely extended by linearity to a normal
state on the algebra $\mathcal{B}\otimes \mathcal{A}$ generated by all
linear combinations $C=\sum \lambda _jB_j\otimes A_j$ due to $\varpi \left(
I\otimes I\right) =\mathrm{tr}\kappa ^{\dagger }\kappa =1$ and 
\begin{eqnarray*}
\varpi \left( C^{\dagger }C\right) &=&\sum_{i,k}\bar{\lambda}_i\lambda _k%
\mathrm{tr}_{\mathcal{G}}\tilde{B}_k\tilde{B}_i^{\dagger }\kappa ^{\dagger
}A_i^{\dagger }A_k\kappa \\
&=&\sum_{i,k}\bar{\lambda}_i\lambda _k\mathrm{tr}_{\mathcal{G}}\tilde{B}
_i^{\dagger }\kappa ^{\dagger }A_i^{\dagger }A_k\kappa \tilde{B}_k=\mathrm{tr%
}_{\mathcal{G}}\chi ^{\dagger }\chi \geq 0,
\end{eqnarray*}
where $\chi =\sum_jA_j\kappa \tilde{B}_j$. This state is pure on $\mathcal{L}
\left( \mathcal{G}\otimes \mathcal{H}\right) $ as it is given by an
amplitude $\vartheta \in \mathcal{G}\otimes \mathcal{H}$ defined as 
\begin{equation*}
\left( \zeta \otimes \eta \right) ^{\dagger }\vartheta =\eta ^{\dagger
}\kappa J\zeta ,\quad \forall \zeta \in \mathcal{G},\eta \in \mathcal{H},
\end{equation*}
and it has the states $\varrho $ and $\varsigma $ as the marginals of $%
\varpi $: 
\begin{equation}
\varpi \left( I\otimes A\right) =\mathrm{tr}_{\mathcal{H}}A\rho ,\quad
\varpi \left( B\otimes I\right) =\mathrm{tr}_{\mathcal{G}}B\sigma .
\label{1.2}
\end{equation}
As follows from the next theorem for the case $\mathcal{F}=\mathbb{C}$ , any
pure state 
\begin{equation*}
\varpi \left( B\otimes A\right) =\vartheta ^{\dagger }\left( B\otimes
A\right) \vartheta ,\quad B\in \mathcal{B},A\in \mathcal{A}
\end{equation*}
given on $\mathcal{L}\left( \mathcal{G}\otimes \mathcal{H}\right) $ by an
amplitude $\vartheta \in \mathcal{G}\otimes \mathcal{H}$ with $\vartheta
^{\dagger }\vartheta =1$, can be achieved by a unique entanglement of its
marginal states $\varsigma $ and $\varrho $.\smallskip

\bigskip

\noindent
{\bf Theorem 2.1.} 
{\it
Let $\varpi :\mathcal{B}\otimes \mathcal{A}\rightarrow \mathbb{C}$ be a compound state 
\begin{equation}
\varpi \left( B\otimes A\right) =\mathrm{tr}_{\mathcal{F}}\upsilon ^{\dagger
}\left( B\otimes A\right) \upsilon ,  \label{1.3}
\end{equation}
defined by an amplitude operator $\upsilon :\mathcal{F}\rightarrow \mathcal{G%
}\otimes \mathcal{H}$ on a separable Hilbert space $\mathcal{F}$ into the
tensor product Hilbert space $\mathcal{G}\otimes \mathcal{H}$ with $\mathrm{%
tr}\upsilon ^{\dagger }\upsilon =1$. Then this state can be achieved as an
entanglement 
\begin{equation}
\varpi \left( B\otimes A\right) =\mathrm{tr}_{\mathcal{G}}\tilde{B}\kappa
^{\dagger }\left( I\otimes A\right) \kappa =\mathrm{tr}_{\mathcal{F}\otimes 
\mathcal{H}}\left( I\otimes A\right) \kappa \tilde{B}\kappa ^{\dagger }
\label{1.4}
\end{equation}
of the states (\ref{1.2}) with $\sigma =\kappa ^{\dagger }\kappa $ and $\rho
=\mathrm{tr}_{\mathcal{F}}\kappa \kappa ^{\dagger }$, where $\kappa $ is an
amplitude operator $\mathcal{G}\rightarrow \mathcal{F}\otimes \mathcal{H}$.
The entangling operator $\kappa $ is uniquely defined by $\tilde{\kappa}%
U=\upsilon $ up to a unitary transformation $U$ of the minimal domain $%
\mathcal{F}=\mathrm{dom}\upsilon $.
}

\bigskip

Note that the entangled state (\ref{1.4}) is written as 
\begin{equation}
\varpi \left( B\otimes A\right) =\mathrm{tr}_{\mathcal{G}}\tilde{B}\pi
\left( A\right) =\mathrm{tr}_{\mathcal{H}}A\pi _{*}\left( \tilde{B}\right) ,
\label{1.6}
\end{equation}
where $\pi \left( A\right) =\kappa ^{\dagger }\left( I\otimes A\right)
\kappa $, bounded by $\left\| A\right\| \sigma \in \mathcal{B}_{*}$ for any $%
A\in \mathcal{L}\left( \mathcal{H}\right) $, is in the predual space $%
\mathcal{B}_{*}\subset \mathcal{B}$ of all trace-class operators in $%
\mathcal{G}$, and $\pi _{*}\left( B\right) =\mathrm{tr}_{\mathcal{F}}\kappa
B\kappa ^{\dagger }$, bounded by $\left\| B\right\| \rho \in \mathcal{A}_{*}$
, is in $\mathcal{A}_{*}\subset \mathcal{A}$. The map $\pi $ is the
Steinspring form \cite{Sti55} of the general completely positive map $%
\mathcal{A}\rightarrow \mathcal{B}_{*}$, written in the eigen-basis $\left\{
\left| k\right\rangle \right\} \subset \mathcal{F}$ of the density operator $\upsilon ^{\dagger }\upsilon $ as 
\begin{equation}
\pi \left( A\right) =\sum_{m,n}\left| m\right\rangle \kappa _m^{\dagger
}\left( I\otimes A\right) \kappa _n\left\langle n\right| ,\quad A\in 
\mathcal{A}  \label{1.7}
\end{equation}
while the dual operation $\pi _{*}$ is the Kraus form \cite{Kra71} of the
general completely positive map $\mathrm{A} \rightarrow \mathcal{A}_{*}$,
given in this basis as 
\begin{equation}
\pi _{*}\left( B\right) =\sum_{n,m}\left\langle n\right| B\left|
m\right\rangle \mathrm{tr}_{\mathcal{F}}\kappa _n\kappa _m^{\dagger }=%
\mathrm{tr}_{\mathcal{G}}\tilde{B}\omega .  \label{1.8}
\end{equation}
It corresponds to the general form 
\begin{equation}
\omega =\sum_{m,n}|n\rangle \langle m|\otimes \mathrm{tr}_{\mathcal{F}
}\kappa _n\kappa _m^{\dagger }  \label{1.9}
\end{equation}
of the density operator $\omega =\upsilon \upsilon ^{\dagger }$ for the
entangled state $\varpi \left( B\otimes A\right) =\mathrm{tr}\left( B\otimes
A\right) \omega $ in this basis, characterized by the weak orthogonality
property 
\begin{equation}
\mathrm{tr}_{\mathcal{F}}\psi \left( m\right) ^{\dagger }\psi \left(
n\right) =\mu \left( n\right) \delta _n^m  \label{1.10}
\end{equation}
in terms of the amplitude operators $\psi \left( n\right) =\left( I\otimes
\langle n|\right) \tilde{\kappa}=\tilde{\kappa}_n$.

\bigskip

\noindent
{\bf Definition 2.1.} 
{\it
The dual map $\pi _{*}:\mathcal{B}\rightarrow \mathcal{A}_{*}$ to a
completely positive map $\pi :\mathcal{A}\rightarrow \mathcal{B}_{*}$,
normalized as $\mathrm{tr}_{\mathcal{G}}\pi \left( I\right) =1$, is called
the quantum entanglement of the state $\varsigma =\pi \left( I\right) $ on $%
\mathcal{B}$ to the state $\varrho =\pi _{*}\left( I\right) $ on $\mathcal{A}
$. The entanglement by 
\begin{equation}
\pi _{*}^{\circ }\left( A\right) =\rho ^{1/2}A\rho ^{1/2}=\pi ^{\circ
}\left( A\right)  \label{1.5}
\end{equation}
of the state $\varsigma =\varrho $ on the algebra $\mathcal{B}=\mathcal{A}$
is called standard for the system $\left( \mathcal{A},\varrho \right) $.
}

\bigskip

The standard entanglement defines the standard compound state 
\begin{equation*}
\varpi _{0}\left( B\otimes A\right) =\mathrm{tr}_{\mathcal{H}}\tilde{B}\rho
^{1/2}A\rho ^{1/2}=\mathrm{tr}_{\mathcal{H}}A\rho ^{1/2}\tilde{B}\rho ^{1/2}
\end{equation*}
on the algebra $\mathcal{A}\otimes \mathcal{A}$, which is pure, given by the
amplitude $\vartheta _{0}$ associated with $\varpi _{0}$ is $\tilde{\kappa}%
_{0}$, where $\kappa _{0}=\rho ^{1/2}$.

\bigskip

\noindent
{\bf Example 2.1.} 
{\it 
In quantum physics the entangled states are usually obtained by a unitary
transformation $U$ of an initial disentangled state, described by the
density operator $\sigma _{0}\otimes \rho _{0}\otimes \tau _{0}$ on the
tensor product Hilbert space $\mathcal{G}\otimes \mathcal{H}\otimes \mathcal{K}$ , that is, 
\begin{equation*}
\varpi \left( B\otimes A\right) =\mathrm{tr}U^{\dagger }\left( B\otimes
A\otimes I\right) U\left( \sigma _{0}\otimes \rho _{0}\otimes \tau
_{0}\right) .
\end{equation*}
In the simple case, when $\mathcal{K}=\mathbb{C}$, $\tau _{0}=1$, the joint
amplitude operator $\upsilon $ is defined on the tensor product $\mathcal{F}=\mathcal{G}\otimes \mathcal{H}_{0}$ with $\mathcal{H}_{0}=\mathrm{ran}\rho_{0}$ as $\upsilon =U_{1}\left( \sigma _{0}\otimes \rho _{0}\right) ^{1/2}$.
The entangling operator $\kappa $, describing the entangled state $\varpi $,
is constructed as it was done in the proof of Theorem 2.1 by transponation of the operator $\upsilon U^{\dagger }$, where $U$ is arbitrary isometric
operator $\mathcal{F}\rightarrow \mathcal{G}\otimes \mathcal{H}_{0}$. The
dynamical procedure of such entanglement in terms of the completely positive
map $\pi _{*}:\mathcal{A}\rightarrow \mathcal{B}_{*}$ is the subject of
Belavkin quantum filtering theory \cite{Bel97}. The quantum filtering
dilation theorem \cite{Bel97} proves that any entanglement $\pi $ can be
obtained the unitary entanglement as the result of quantum filtering by
tracing out some degrees of freedom of a quantum environment, described by the density operator $\tau _{0}$ on the Hilbert space $\mathcal{K}$, even in the continuous time case.
}

\bigskip

\section{C- and D-Entanglements and Encodings}

The compound states play the role of joint input-output probability measures
in classical information channels, and can be pure in quantum case even if
the marginal states are mixed. The pure compound states achieved by an
entanglement of mixed input and output states exhibit new, non-classical
type of correlations which are responsible for the EPR type paradoxes in the
interpretation of quantum theory. The mixed compound states on $\mathcal{B}%
\otimes \mathcal{A}$ which are given as the convex combinations 
\begin{equation*}
\varpi =\sum_{n}\varsigma _{n}\otimes \varrho _{n}\mu \left( n\right) ,\quad
\mu \left( n\right) \geq 0,\;\sum_{n}\mu \left( n\right) =1
\end{equation*}
of tensor products of pure or mixed normalized states $\varrho _{n}\in 
\mathcal{A}_{*}$, $\varsigma _{n}\in \mathcal{B}_{*}$ as in classical case,
do not exhibit such paradoxical behavior, and are usually considered as the
proper candidates for the input-output states in the communication channels.
Such separable compound states are achieved by c-entanglements, the convex
combinations of the primitive entanglements $B\mapsto \mathrm{tr}_{\mathcal{G%
}}B\omega _{n}$, given by the density operators $\omega _{n}=\sigma
_{n}\otimes \rho _{n}$ of the product states $\varpi _{n}=\varsigma
_{n}\otimes \varrho _{n}$: 
\begin{equation}
\pi _{*}\left( B\right) =\sum_{n}\rho _{n}\mathrm{tr}_{\mathcal{G}}B\sigma
_{n}\mu \left( n\right) ,  \label{2.1}
\end{equation}
A compound state of this sort was introduced by Ohya \cite{Ohy83,Ohy89} in
order to define the quantum mutual entropy expressing the amount of
information transmitted from an input quantum system to an output quantum
system through a quantum channel, using a Schatten decomposition $\sigma
=\sum_{n}\sigma _{n}\mu \left( n\right) $, $\sigma _{n}=|n\rangle \langle n|$
of the input density operator $\sigma $. It corresponds to a particular,
diagonal type 
\begin{equation}
\pi \left( A\right) =\sum_{n}|n\rangle \kappa _{n}^{\dagger }\left( I\otimes
A\right) \kappa _{n}\langle n|  \label{2.4}
\end{equation}
of the entangling map (\ref{1.7}) in an eigen-basis $\left\{ |n\rangle
\right\} \in \mathcal{G}$ of the density operator $\sigma $, and is
discussed in this section.

Let us consider a finite or infinite input system indexed by the natural
numbers $n\in \mathbf{N}$. The associated space $\mathcal{G}\subseteq
l^2\left( \mathbf{N}\right) $ is the Hilbert space of the input system
described by a quantum projection-valued measure $n\mapsto |n\rangle \langle
n|$ on $\mathbb{N}$, given an orthogonal partition of unity $I=\sum
|n\rangle \langle n|$ $\in \mathcal{B}$ of the finite or infinite
dimensional input Hilbert space $\mathcal{G}$. Each input pure state,
identified with the one-dimensional density operator $|n\rangle \langle
n|\in \mathcal{B}$ corresponding to the elementary symbol $n\in \mathbb{N}$,
defines the elementary output state $\varrho _n$ on $\mathcal{A}$. If the
elementary states $\varrho _n$ are pure, they are described by output
amplitudes $\eta _n\in \mathcal{H}$ satisfying $\eta _n^{\dagger }\eta _n=1=%
\mathrm{tr}\rho _n $, where $\rho _n=$ $\eta _n\eta _n^{\dagger }$ are the
corresponding output one-dimensional density operators. If these amplitudes
are non-orthogonal $\eta _m^{\dagger }\eta _n\neq \delta _n^m$, they cannot
be identified with the input amplitudes $|n\rangle $.

The elementary joint input-output states are given by the density operators $%
|n\rangle \langle n|\otimes \rho _n$ in $\mathcal{G}\otimes \mathcal{H}$.
Their mixtures 
\begin{equation}
\omega =\sum_n\mu \left( n\right) |n\rangle \langle n|\otimes \rho _n,
\label{2.2}
\end{equation}
define the compound states on $\mathcal{B}\otimes \mathcal{A}$, given by the
quantum correspondences $n\mapsto |n\rangle \langle n|$ with the
probabilities $\mu \left( n\right) $. Here we note that the quantum
correspondence is described by a classical-quantum channel, and the general
d-compound state for a quantum-quantum channel in quantum communication can
be obtained in this way due to the orthogonality of the decomposition (\ref
{2.2}), corresponding to the orthogonality of the Schatten decomposition $%
\sigma =\sum_n|n\rangle \mu \left( n\right) \langle n|$ for $\sigma =\mathrm{%
\ tr}_{\mathcal{H}}\omega $.

The comparison of the general compound state (\ref{1.9}) with (\ref{2.2})
suggests that the quantum correspondences are described as the diagonal
entanglements 
\begin{equation}
\pi _{*}\left( B\right) =\sum_{n}\mu \left( n\right) \langle n|B|n\rangle
\rho _{n},  \label{2.3}
\end{equation}
They are dual to the orthogonal decompositions (\ref{2.4}): 
\begin{equation*}
\pi \left( A\right) =\sum_{n}\mu \left( n\right) |n\rangle \eta
_{n}^{\dagger }A\eta _{n}\langle n|=\sum_{n}|n\rangle \eta \left( n\right)
^{\dagger }A\eta \left( n\right) \langle n|,
\end{equation*}
\newline
where $\eta \left( n\right) =\mu \left( n\right) ^{1/2}\eta _{n}$. These are
the entanglements with the stronger orthogonality 
\begin{equation}
\psi \left( m\right) \psi \left( n\right) ^{\dagger }=\mu \left( n\right)
\delta _{n}^{m},  \label{2.5}
\end{equation}
for the amplitude operators $\psi \left( n\right) :\mathcal{F}\rightarrow 
\mathcal{H}$ of the decomposition of the amplitude operator $\upsilon
=\sum_{n}|n\rangle \otimes \psi \left( n\right) $ in comparison with the
orthogonality (\ref{1.10}). The orthogonality (\ref{2.5}) can be achieved in
the following manner: Take in (\ref{1.7}) $\kappa _{n}=|n\rangle \otimes
\eta \left( n\right) $ with $\langle m|n\rangle =\delta _{n}^{m}$ so that 
\begin{equation*}
\kappa _{m}^{\dagger }\left( I\otimes A\right) \kappa _{n}=\mu \left(
n\right) \eta _{n}^{\dagger }A\eta _{n}\delta _{n}^{m}
\end{equation*}
for any $A\in \mathcal{A}$. Then the strong orthogonality condition (\ref
{2.5}) is fulfilled by the amplitude operators $\psi \left( n\right) =\eta
\left( n\right) \langle n|=\tilde{\kappa}_{n}$, and 
\begin{equation*}
\kappa ^{\dagger }\kappa =\sum_{n}\mu \left( n\right) |n\rangle \langle
n|=\sigma ,\;\kappa \kappa ^{\dagger }=\sum_{n}\eta \left( n\right) \eta
\left( n\right) ^{\dagger }=\rho .
\end{equation*}
It corresponds to the amplitude operator for the compound state (\ref{2.2})
of the form 
\begin{equation}
\upsilon =\sum_{n}\left| n\right\rangle \otimes \psi \left( n\right) U,
\label{2.6}
\end{equation}
where $U$ is arbitrary unitary operator from $\mathcal{F}$ onto $\mathcal{G}$
, i.e. $\upsilon $ is unitary equivalent to the diagonal amplitude operator 
\begin{equation*}
\kappa =\sum_{n}|n\rangle \langle n|\otimes \eta \left( n\right)
\end{equation*}
on $\mathcal{F}=\mathcal{G}$ into $\mathcal{G}\otimes \mathcal{H}$. Thus, we
have proved the following theorem in the case of pure output states $\rho
_{n}=\eta _{n}\eta _{n}^{\dagger }$.

\bigskip

\noindent
{\bf Theorem 3.1.} 
{\it
Let $\pi $ be the operator (\ref{2.2}), defining a d-compound state of the
form 
\begin{equation}
\varpi \left( B\otimes A\right) =\sum_{n}\langle n|B|n\rangle \mathrm{tr}_{%
\mathcal{F}_{n}}\psi _{n}^{\dagger }A\psi _{n}\mu \left( n\right)
\label{2.7}
\end{equation}
Then it corresponds to the entanglement by the orthogonal decomposition (\ref
{2.4}) mapping the algebra $\mathcal{A}$ into a diagonal subalgebra of $%
\mathcal{B}$.
}

\bigskip

Note that (2.9) defines the general form of a positive map on $\mathcal{A}$ with values in the simultaneously diagonal trace-class operators in 
$\mathrm{A}$.

\bigskip

\noindent
{\bf Definition 3.1.} 
{\it
The completely positive convex combination (\ref{2.1}) is called
c-entanglement, and is called d-entanglement, or quantum encoding if it has
the diagonal form (\ref{2.3}) on $\mathcal{B}$. The d-entanglement is called
o-entanglement and compound state is called o-compound if all density
operators $\rho _{n}$ are orthogonal: $\rho _{m}\rho _{n}=\rho _{n}\rho _{m}$
for all $m$ and $n$.
}

\bigskip

Note that due to the commutativity of the operators $B\otimes I$ with $
I\otimes A$ on $\mathcal{G}\otimes \mathcal{H}$, one can treat the
correspondences as the nondemolition measurements \cite{Bel94} in $\mathcal{B}$ with respect to $\mathcal{A}$. So, the compound state is the state
prepared for such measurements on the input $\mathcal{G}$. It coincides with
the mixture of the states, corresponding to those after the measurement
without reading the sent message. The set of all d-entanglements
corresponding to a given Schatten decomposition of the input state $\sigma $
on $\mathcal{B}$ is obviously convex with the extreme points given by the
pure output states $\rho _n$ on $\mathcal{A}$, corresponding to a not
necessarily orthogonal decompositions $\rho =\sum_n\rho \left( n\right) $
into one-dimensional density operators $\rho \left( n\right) =\mu \left(
n\right) \rho _n.$

The Schatten decompositions $\rho =\sum_n\lambda \left( n\right) \rho _n$
correspond to the extreme d-entanglements, $\rho _n=\eta _n\eta _n^{\dagger
} $, $\mu \left( n\right) =\lambda \left( n\right) $, characterized by
orthogonality $\rho _m\rho _n=0$, $m\neq n$ . They form a convex set of
d-entanglements with mixed commuting $\rho _n$ for each Schatten
decomposition of $\rho $. The orthogonal d-entanglements were used in \cite
{AOW96} to construct a particular type of Accardi's transitional
expectations \cite{Acc74} and to define the entropy in a quantum dynamical
system via such transitional expectations.

The established structure of the general q-compound states suggests also the
general form 
\begin{equation*}
\Phi _{\ast }\left( B,\varrho _{0}\right) =\mathrm{tr}_{\mathcal{F}%
_{1}}X^{\dagger }\left( B\otimes \rho _{0}\right) X=\mathrm{tr}_{\mathcal{G}%
}\left( \tilde{B}\otimes I\right) Y\left( I\otimes \rho _{0}\right)
Y^{\dagger }
\end{equation*}
of transitional expectations $\Phi _{\ast }:\mathcal{B}\times \mathcal{A}%
_{\ast }^{\circ }\rightarrow \mathcal{A}_{\ast }$, describing the
entanglements $\pi _{\ast }=\Phi _{\ast }\left( \varrho _{0}\right) $ of the
states $\varsigma =\pi \left( I\right) $ to $\varrho =\pi _{\ast }\left(
I\right) $ for each initial state $\varrho _{0}\in \mathcal{A}_{\ast
}^{\circ }$ with the density operator $\rho _{0}\in \mathcal{A}^{\circ
}\subseteq \mathcal{L}\left( \mathcal{H}_{0}\right) $ by $\pi _{\ast }\left(
B\right) =\mathrm{tr}_{\mathcal{F}}\kappa \left( B\otimes I\right) \kappa
^{\dagger }$, where $\kappa =X^{\dagger }\left( I\otimes \rho _{0}\right)
^{1/2}$. It is given by an entangling transition operator $X:\mathcal{F}%
\otimes \mathcal{H}\rightarrow \mathcal{G}\otimes \mathcal{H}_{0}$, which is
defined by a transitional amplitude operator $Y:\mathcal{H}_{0}\otimes 
\mathcal{F}\rightarrow \mathcal{G}\otimes \mathcal{H}$ up to a unitary
operator $U$ in $\mathcal{F}$ as 
\begin{equation*}
\left( \zeta \otimes \eta _{0}\right) ^{\dagger }X\left( U\xi \otimes \eta
\right) =\left( \eta _{0}\otimes J\xi \right) ^{\dagger }Y^{\dagger }\left(
J\zeta \otimes \eta \right) \text{.}
\end{equation*}
The dual map $\Phi :\mathcal{A}\rightarrow \mathcal{B}_{\ast }\otimes 
\mathcal{A}^{\circ }$ is obviously normal and completely positive, 
\begin{equation}
\Phi \left( A\right) =X\left( I\otimes A\right) X^{\dagger }\in \mathcal{B}%
_{\ast }\otimes \mathcal{A}^{\circ },\;\forall A\in \mathcal{A},  \label{2.9}
\end{equation}
with $\mathrm{tr}_{\mathcal{G}}\Phi \left( I\right) =I^{\circ }$, and is
called filtering map with the output states 
\begin{equation*}
\varsigma =\mathrm{tr}_{\mathcal{H}_{0}}\Phi \left( I\right) \left( I\otimes
\rho _{0}\right)
\end{equation*}
in the theory of CP flows \cite{Bel97} over $\mathcal{A}=\mathcal{A}^{\circ
} $. The operators $Y$ normalized as $\mathrm{tr}_{\mathcal{F}}Y^{\dagger
}Y=I^{\circ }$ describe $\mathcal{A}$-valued q-compound states 
\begin{equation*}
\mathrm{E}\left( B\otimes A\right) =\mathrm{tr}_{\mathcal{F}}Y^{\dagger
}\left( B\otimes A\right) Y=\mathrm{tr}_{\mathcal{G}}\left( \tilde{B}\otimes
I\right) \Phi \left( A\right) ,
\end{equation*}
defined as the normal completely positive maps $\mathcal{B}\otimes \mathcal{A%
}\rightarrow \mathcal{A}^{\circ }$ with $\mathrm{E}\left( I\otimes I\right)
=I^{\circ }$ .

If the $\mathcal{A}$-valued compound state has the diagonal form given by
the orthogonal decomposition 
\begin{equation}
\Phi \left( A\right) =\sum_n|n\rangle \mathrm{tr}_{\mathcal{F}}\Psi \left(
n\right) ^{\dagger }A\Psi \left( n\right) \langle n|,  \label{2.10}
\end{equation}
corresponding to $Y$ $=\sum_n|n\rangle \otimes \Psi \left( n\right) $, where 
$\Psi \left( n\right) :\mathcal{H}_0\otimes \mathcal{F}\rightarrow \mathcal{H%
}$, it is achieved by the d-transitional expectations 
\begin{equation*}
\Phi _{*}\left( B,\varrho _0\right) =\sum_n\langle n|B|n\rangle \Psi \left(
n\right) \left( \rho _0\otimes I\right) \Psi \left( n\right) ^{\dagger }.
\end{equation*}
The d-transitional expectations correspond to the instruments \cite{DaL71}
of the dynamical theory of quantum measurements. The elementary filters 
\begin{equation*}
\Theta _n\left( A\right) =\frac 1{\mu \left( n\right) }\mathrm{tr}_{\mathcal{%
\ F}}\Psi ^{\dagger }\left( n\right) A\Psi \left( n\right) ,\quad \mu \left(
n\right) =\mathrm{tr}\Psi \left( n\right) \left( \rho _0\otimes I\right)
\Psi ^{\dagger }\left( n\right)
\end{equation*}
define posterior states $\varrho _n=\varrho _0\Theta _n$ on $\mathcal{A}$
for quantum nondemolition measurements in $\mathcal{B}$, which are called
indirect if the corresponding density operators $\rho _n$ are
non-orthogonal. They describe the posterior states with orthogonal 
\begin{equation*}
\rho _n=\Psi _n\left( \rho _0\otimes I\right) \Psi _n^{\dagger },\quad \Psi
_n=\Psi \left( n\right) /\mu \left( n\right) ^{1/2}
\end{equation*}
for all $\rho _0$ iff $\Psi \left( n\right) ^{\dagger }\Psi \left( n\right)
=\delta _n^mM\left( n\right) $.

\section{Quantum Entropy via Entanglements}

As it was shown in the previous section, the diagonal entanglements describe
the classical-quantum encodings $\varkappa :\mathcal{B}\rightarrow \mathcal{A%
}_{*}$, i.e. correspondences of classical symbols to quantum, in general not
orthogonal and pure, states. As we have seen in contrast to the classical
case, not every entanglement can be achieved in this way. The general
entangled states $\varpi $ are described by the density operators $\omega
=\upsilon \upsilon ^{\dagger }$ of the form (\ref{1.9}) which are not
necessarily block-diagonal in the eigen-representation of the density
operator $\sigma $, and they cannot be achieved even by a more general
c-entanglement (\ref{2.1}). Such nonseparable entangled states are called in 
\cite{Ohy89} the quasicompound (q-compound) states, so we can call also the
quantum nonseparable correspondences the quasi-encodings (q-encodings) in
contrast to the d-correspondences, described by the diagonal entanglements.

As we shall prove in this section, the most informative for a quantum system 
$\left( \mathcal{A},\varrho \right) $ is the standard entanglement $\pi
_{*}^{\circ }=\pi _0$ of the probe system $\left( \mathcal{B}^{\circ
},\varsigma _0\right) =\left( \mathcal{A},\varrho \right) $, described in (%
\ref{1.5}). The other extreme cases of the self-dual input entanglements 
\begin{equation*}
\pi _{*}\left( A\right) =\sum_n\rho \left( n\right) ^{1/2}A\rho \left(
n\right) ^{1/2}=\pi \left( A\right) ,
\end{equation*}
are the pure c-entanglements, given by the decompositions $\rho =\sum \rho
\left( n\right) $ into pure states $\rho \left( n\right) =\eta _n\eta
_n^{\dagger }\mu \left( n\right) $. We shall see that these c-entanglements,
corresponding to the separable states 
\begin{equation}
\omega =\sum_n\eta _n\eta _n^{\dagger }\otimes \eta _n\eta _n^{\dagger }\mu
\left( n\right) ,  \label{4.0}
\end{equation}
are in general less informative then the pure d-entanglements, given in an
orthonormal basis $\left\{ \eta _n^{\circ }\right\} \subset \mathcal{H}$ by 
\begin{equation*}
\pi ^{\circ }\left( A\right) =\sum_n\eta _n^{\circ }\eta _n^{\dagger }A\eta
_n\eta _n^{\circ \dagger }\mu \left( n\right) \neq \pi _{*}^{\circ }\left(
A\right) .
\end{equation*}

Now, let us consider the entangled mutual entropy and quantum entropies of
states by means of the above three types of compound states. To define the
quantum mutual entropy, we need the relative entropy \cite{Lin73, Ara76,Ume}
of the compound state $\varpi $ with respect to a reference state $\varphi $
on the algebra $\mathcal{A}\otimes \mathcal{B}$. It is defined by the
density operators $\omega ,\phi \in \mathcal{B}\otimes \mathcal{A}$ of these
states as 
\begin{equation}
S\left( \varpi ,\varphi \right) =\mathrm{tr}\omega \left( \ln \omega -\ln
\phi \right) .  \label{4.1}
\end{equation}
It has a positive value $S\left( \varpi ,\varphi \right) \in [0,\infty ]$ if
the states are equally normalized, say (as usually) $\mathrm{tr}\omega =1=%
\mathrm{tr}\phi $, and it can be finite only if the state $\varpi $ is
absolutely continuous with respect to the reference state $\varphi $, i.e.
iff $\varpi \left( E\right) =0$ for the maximal null-orthoprojector $E\phi
=0 $.

The mutual entropy $I_{\omega }\left( \mathcal{A},\mathcal{B}\right) $ of a
compound state $\varpi $ achieved by an entanglement $\pi _{*}:$ $\mathcal{B}
\rightarrow \mathcal{A}_{*}$ with the marginals 
\begin{equation*}
\varsigma \left( B\right) =\varpi \left( B\otimes I\right) =\mathrm{tr}_{%
\mathcal{G}}B\sigma ,\;\varrho \left( A\right) =\varpi \left( I\otimes
A\right) =\mathrm{tr}_{\mathcal{H}}A\rho
\end{equation*}
is defined as the relative entropy (\ref{4.1}) with respect to the product
state $\varphi =\varsigma \otimes \varrho $: 
\begin{equation}
I_{\mathcal{A},\mathcal{B}}\left( \varpi \right) =\mathrm{tr}\omega \left(
\ln \omega -\ln \left( \sigma \otimes I\right) -\ln \left( I\otimes \rho
\right) \right) .  \label{4.3}
\end{equation}
Here the operator $\omega $ is uniquely defined by the entanglement $\pi
_{*} $ as its density in (\ref{1.8}), or the $\mathcal{G}$-transposed to the
operator $\tilde{\omega}$ in 
\begin{equation*}
\pi \left( A\right) =\kappa ^{\dagger }\left( I\otimes A\right) \kappa =%
\mathrm{tr}_{\mathcal{H}}A\tilde{\omega}.
\end{equation*}
This quantity describes an information gain in a quantum system $\left( 
\mathcal{A},\varrho \right) $ via an entanglement $\pi _{*}$ of another
system $\left( \mathcal{B},\varsigma \right) .$ It is naturally treated as a
measure of the strength of an entanglement, having zero value only for
completely disentangled states, corresponding to $\varpi =\varsigma \otimes
\varrho $.

The following proposition follows from the monotonicity property \cite
{Uhl,OP93} 
\begin{equation}
\varpi =\mathrm{K}_{*}\varpi _{0},\varphi =\mathrm{K}_{*}\varphi
_{0}\Rightarrow S\left( \varpi ,\varphi \right) \leq S\left( \varpi
_{0},\varphi _{0}\right) .  \label{4.4}
\end{equation}
of the general relative entropy on a von Neuman algebra $\mathcal{M}$ with
respect to the predual $\mathrm{K}_{*}$ to any normal completely positive
unital map $\mathrm{K}:\mathcal{M}\rightarrow \mathcal{M}^{\circ }$.

\bigskip

\noindent
{\bf Proposition 4.1.} 
{\it
Let $\pi _{*}^{\circ }:\mathcal{B}^{\circ }\rightarrow \mathcal{A}_{*}$ be
an entanglement $\pi _{*}^{\circ }$ of a state $\varsigma _{0}=\pi ^{\circ
}\left( I\right) $ on a discrete decomposable algebra $\mathcal{B}^{\circ
}\subseteq \mathcal{L}\left( \mathcal{G}_{0}\right) $ to the state $\varrho
=\pi _{*}^{\circ }\left( I\right) $ on $\mathcal{A}$, and $\pi _{*}=\pi
_{*}^{\circ }\mathrm{K}$ be an entanglement defined as the composition with
a normal completely positive unital map $\mathrm{K}:\mathcal{B}\rightarrow 
\mathcal{B}^{\circ }$. Then $I_{\mathcal{A},\mathcal{B}}\left( \varpi
\right) \leq I_{\mathcal{A},\mathcal{B}^{\circ }}\left( \varpi _{0}\right) $
, where $\varpi ,\varpi _{0}$ are the compound states achieved by $\pi
_{*}^{\circ }$ , $\pi _{*}$ respectively. In particular, for any
c-entanglement $\pi _{*}$ to $\left( \mathcal{A},\varsigma \right) $ there
exists a not less informative d-entanglement $\pi _{*}^{\circ }=\varkappa $
with an Abelian $\mathcal{B}^{\circ }$, and the standard entanglement $\pi
_{0}\left( A\right) =\rho ^{1/2}A\rho ^{1/2}$ of $\varsigma _{0}=\varrho $
on $\mathcal{B}^{\circ }=\mathcal{A}$ is the maximal one in this sense.
}

\bigskip

Note that any extreme d-entanglement 
\begin{equation*}
\pi _{*}^{\circ }\left( B\right) =\sum_{n}\langle n|B|n\rangle \rho
_{n}^{\circ }\mu \left( n\right) ,\;B\in \mathcal{B}^{\circ },
\end{equation*}
with $\rho =\sum_{n}\rho _{n}^{\circ }\mu \left( n\right) $ decomposed into
pure normalized states $\rho _{n}^{\circ }=\eta _{n}\eta _{n}^{\dagger }$,
is maximal among all c-entanglements in the sense $I_{\mathcal{A},\mathcal{B}%
}\left( \varpi _{0}\right) \geq I_{\mathcal{A},\mathcal{B}}\left( \varpi
\right) $. This is because $\mathrm{tr}\rho _{n}^{\circ }\ln \rho
_{n}^{\circ }=0$, and therefore the information gain 
\begin{equation*}
I_{\mathcal{A},\mathcal{B}}\left( \varpi \right) =\sum_{n}\mu \left(
n\right) \mathrm{tr}\rho _{n}\left( \ln \rho _{n}-\ln \rho \right) .
\end{equation*}
with a fixed $\pi _{*}\left( I\right) =\rho $ achieves its supremum $-%
\mathrm{tr}_{\mathcal{H}}\rho \ln \rho $ at any such extreme d-entanglement $%
\pi _{*}^{\circ }$. Thus the supremum of the information gain (\ref{4.3})
over all c-entanglements to the system $\left( \mathcal{A},\varrho \right) $
is the von Neumann entropy 
\begin{equation}
S_{\mathcal{A}}\left( \varrho \right) =-\mathrm{tr}_{\mathcal{H}}\rho \ln
\rho .  \label{4.5}
\end{equation}
It is achieved on any extreme $\pi _{*}^{\circ }$, for example given by the
maximal Abelian subalgebra $\mathcal{B}^{\circ }\subseteq \mathcal{A}$, with
the measure $\mu =\lambda $, corresponding to a Schatten decomposition $\rho
=\sum_{n}\eta _{n}^{\circ }\eta _{n}^{\circ \dagger }\lambda \left( n\right) 
$, $\eta _{m}^{\circ \dagger }\eta _{n}^{\circ }=\delta _{n}^{m}$. The
maximal value $\ln \,\mathrm{rank}\mathcal{A}$ of the von Neumann entropy is
defined by the dimensionality $\mathrm{rank}\mathcal{A}=\dim \mathcal{B}%
^{\circ }$ of the maximal Abelian subalgebra of the decomposable algebra $%
\mathcal{A}$, i.e. by $\dim \mathcal{H}$.

\noindent
{\bf Definition 4.1.} 
{\it
The maximal mutual entropy 
\begin{equation}
\widetilde{S}_{\mathcal{A}}\left( \varrho \right) =\sup_{\pi _{*}\left(
I\right) =\rho }I_{\mathcal{A},\mathcal{B}}\left( \varpi \right) =I_{%
\mathcal{A},\mathcal{B}^{\circ }}\left( \varpi _{0}\right) ,  \label{4.8}
\end{equation}
achieved on $\mathcal{B}^{\circ }=\mathcal{A}$ by the standard
q-entanglement $\pi _{*}^{\circ }\left( A\right) =\rho ^{1/2}A\rho ^{1/2}$
for a fixed state $\varrho \left( A\right) =\mathrm{tr}_{\mathcal{H}}A\rho $
, is called q-entropy of the state $\varrho $. The differences 
\begin{equation*}
\widetilde{S}_{\mathcal{B}|\mathcal{A}}\left( \varpi \right) =\widetilde{S}_{%
\mathcal{B}}\left( \varsigma \right) -I_{\mathcal{A},\mathcal{B}}\left(
\varpi \right)
\end{equation*}
\begin{equation*}
S_{\mathcal{B}|\mathcal{A}}\left( \varpi \right) =S_{\mathcal{B}}\left(
\varsigma \right) -I_{\mathcal{A},\mathcal{B}}\left( \varpi \right)
\end{equation*}
are respectively called the q-conditional entropy on $\mathcal{B}$ with
respect to $\mathcal{A}$ and the degree of disentanglement for the compound
state $\varpi $.
}

\bigskip

Obviously, $\widetilde{S}_{\mathcal{B}|\mathcal{A}}\left( \varpi \right) $
is positive in contrast to the disentanglement $S_{\mathcal{B}|\mathcal{A}
}\left( \varpi \right) $, having the positive maximal value $S_{\mathcal{B}|%
\mathcal{A}}\left( \varpi \right) =S_{\mathcal{B}}\left( \varsigma \right) $
in the case $\varpi =\varsigma \otimes \varrho $ of complete
disentanglement, but which can achieve also a negative value 
\begin{equation}
\inf_{\pi _{*}\left( I\right) =\rho }D_{\mathcal{B}|\mathcal{A}}\left(
\varpi \right) =S_{\mathcal{A}}\left( \varsigma \right) -\widetilde{S}_{%
\mathcal{A}}\left( \varrho \right) =\mathrm{tr}\rho \ln \rho  \label{4.9}
\end{equation}
for the entangled states as the following theorem states. Obviously $S_{%
\mathcal{A}}\left( \varrho \right) =\widetilde{S}_{\mathcal{A}}\left(
\varrho \right) $ if the algebra $\mathcal{A}$ is completely decomposable,
i.e. Abelian, and the maximal value $\ln \,\mathrm{rank}\mathcal{A}$ of $S_{%
\mathcal{A}}\left( \varrho \right) $ can be written as $\ln \dim \mathcal{A}$
in this case. The disentanglement $S_{\mathcal{B}|\mathcal{A}}\left( \varpi
\right) $ is always positive in this case, as well as in the case of Abelian 
$\mathcal{B}$ when $S_{\mathcal{B}|\mathcal{A}}\left( \varpi \right) =%
\widetilde{S}_{\mathcal{B}|\mathcal{A}}\left( \varpi \right) $.

\bigskip

\noindent
{\bf Theorem 4.2.} 
{\it
The q-entropy for the simple algebra $\mathcal{A}=\mathcal{L}\left( \mathcal{H}\right) $ is given by the formula 
\begin{equation}
\widetilde{S}_{\mathcal{A}}\left( \varrho \right) =-2\mathrm{tr}_{\mathcal{H}%
}\rho \ln \rho =2S_{\mathcal{A}}\left( \rho \right) ,  \label{4.10}
\end{equation}
It is positive, $\widetilde{S}_{\mathcal{A}}\left( \varrho \right) \in
[0,\infty ]$, and if $\mathcal{A}$ is finite dimensional, it is bounded,
with the maximal value $\widetilde{S}_{\mathcal{A}}\left( \varrho ^{\circ
}\right) =\ln \dim \mathcal{A}$ which is achieved on the tracial $\rho
^{\circ }=\left( \dim \mathcal{H}\right) ^{-1}I$, where $\dim \mathcal{A}%
=\left( \dim \mathcal{H}\right) ^{2}$ .
}

\bigskip

\section{Quantum Channel and its Q-Capacity}

Let $\mathcal{H}_{0}$ be a Hilbert space describing a quantum input system
and $\mathcal{H}$ describe its output Hilbert space. A quantum channel is an
affine operation sending each input state defined on $\mathcal{H}_{0}$ to an
output state defined on $\mathcal{H}$ such that the mixtures of states are
preserved. A deterministic quantum channel is given by a linear isometry $Y%
\mathrm{:\mathcal{H}}_{0}\rightarrow \mathrm{\mathcal{H}}$ with $Y^{\dagger
}Y=I^{\circ }$ ($I^{\circ }$ is the identify operator in $\mathrm{\mathcal{H}%
}_{0}$) such that each input state vector $\eta \in \mathrm{\mathcal{H}}_{0}$%
, $\left\| \eta \right\| =1$ is transmitted into an output state vector $%
Y\eta \in \mathcal{H}$, $\left\| Y\eta \right\| =1$. The orthogonal mixtures 
$\rho _{0}=\sum_{n}\mu \left( n\right) \rho _{n}^{\circ }$ of the pure input
states $\rho _{n}^{\circ }=\eta _{n}^{\circ }\eta _{n}^{\circ \dagger }$ are
sent into the orthogonal mixtures $\rho =\sum_{n}\mu \left( n\right) \rho
_{n}$ of the corresponding pure states $\rho _{n}=Y\rho _{n}^{\circ
}Y^{\dagger }$.

A noisy quantum channel sends pure input states $\varrho _0$ into mixed ones 
$\varrho =\Lambda ^{*}\left( \varrho _0\right) $ given by the dual $\Lambda
^{*}$ to a normal completely positive unital map $\Lambda :\mathcal{A}
\rightarrow \mathcal{A}_0$, 
\begin{equation*}
\Lambda \left( A\right) =\mathrm{tr}_{\mathcal{F}_1}Y^{\dagger }AY,\mathrm{\
\quad }A\in \mathrm{\mathcal{A}}
\end{equation*}
where $Y$ is a linear operator from $\mathrm{\mathcal{H}}_0\otimes \mathcal{F%
}_{+}$ to $\mathrm{\mathcal{H}}$ with $\mathrm{tr}_{\mathcal{F}
_{+}}Y^{\dagger }Y=I^{\circ }$, and $\mathcal{F}_{+}$ is a separable Hilbert
space of quantum noise in the channel. Each input mixed state $\varrho _0$
on $\mathcal{A}^{\circ }\subseteq \mathcal{L}\left( \mathcal{H}_0\right) $
is transmitted into an output state $\varrho =\varrho _0\Lambda $ given by
the density operator 
\begin{equation*}
\Lambda _{*}\left( \rho _0\right) =Y\left( \rho _0\otimes I^{+}\right)
Y^{\dagger }\in \mathcal{A}_{*}
\end{equation*}
for each density operator $\rho _0\in \mathcal{A}_{*}^{\circ }$, where $%
I^{+} $ is the identity operator in $\mathcal{F}_{+}$. Without loss of
generality we can assume that the input algebra $\mathcal{A}^{\circ }$ is
the smallest decomposable algebra, generated by the range $\Lambda \left( 
\mathcal{A}\right) $ of the given map $\Lambda $.

The input entanglements $\varkappa :\mathcal{B}\rightarrow \mathcal{A}%
_{*}^{\circ }$ described as normal CP maps with $\varkappa \left( I\right)
=\varrho _{0}$, define the quantum correspondences (q-encodings) of probe
systems $\left( \mathcal{B},\varsigma \right) $, $\varsigma =\varkappa
^{*}\left( I\right) $, to $\left( \mathcal{A}^{\circ },\varrho _{0}\right) $%
. As it was proven in the previous section, the most informative is the
standard entanglement $\varkappa =\pi _{*}^{\circ }$, at least in the case
of the trivial channel $\Lambda =\mathrm{I}$. This extreme input
q-entanglement 
\begin{equation*}
\pi ^{\circ }\left( A^{\circ }\right) =\rho _{0}^{1/2}A^{\circ }\rho
_{0}^{1/2}=\pi _{*}^{\circ }\left( A^{\circ }\right) ,\quad A^{\circ }\in 
\mathcal{A}^{\circ },
\end{equation*}
corresponding to the choice $\left( \mathcal{B},\varsigma \right) =\left( 
\mathcal{A}^{\circ },\varrho _{0}\right) $, defines the following density
operator 
\begin{equation}
\omega =\left( \mathrm{I}\otimes \Lambda \right) _{*}\left( \omega
_{q}^{\circ }\right) ,\quad \omega _{q}^{\circ }=\vartheta _{0}\vartheta
_{0}^{\dagger }  \label{3.1}
\end{equation}
of the input-output compound state $\varpi _{q}^{\circ }\Lambda $ on $%
\mathcal{\ A}^{\circ }\otimes \mathcal{A}$. It is given by the amplitude $%
\vartheta _{0}\in \mathcal{H}_{0}^{\otimes 2}$ defined as $\tilde{\vartheta}%
_{0}=\rho _{0}^{1/2}$. The other extreme cases of the self-dual input
entanglements, the pure c-entanglements corresponding to (\ref{4.0}), can be
less informative then the d-entanglements, given by the decompositions $\rho
_{0}=\sum \rho _{0}\left( n\right) $ into pure states $\rho _{0}\left(
n\right) =\eta _{n}\eta _{n}^{\dagger }\mu \left( n\right) $. They define
the density operators 
\begin{equation}
\omega =\left( \mathrm{I}\otimes \Lambda \right) _{*}\left( \omega
_{d}^{\circ }\right) ,\quad \omega _{d}^{\circ }=\sum_{n}\eta _{n}^{\circ
}\eta _{n}^{\circ \dagger }\otimes \eta _{n}\eta _{n}^{\dagger }\mu
_{0}\left( n\right) ,  \label{3.2}
\end{equation}
of the $\mathcal{A}^{\circ }\otimes \mathcal{A}$-compound state $\varpi
_{d}^{\circ }\Lambda $, which are known as the Ohya compound states $\varpi
_{o}^{\circ }\Lambda $ \cite{Ohy83} in the case 
\begin{equation*}
\rho _{0}\left( n\right) =\eta _{n}^{\circ }\eta _{n}^{\circ \dagger
}\lambda _{0}\left( n\right) ,\quad \eta _{m}^{\circ \dagger }\eta
_{n}^{\circ }=\delta _{n}^{m},
\end{equation*}
of orthogonality of the density operators $\rho _{0}\left( n\right) $
normalized to the eigen-values $\lambda _{0}\left( n\right) $ of $\rho _{0}$%
. They are described by the input-output density operators 
\begin{equation}
\omega =\left( \mathrm{I}\otimes \Lambda \right) _{*}\left( \omega
_{o}^{\circ }\right) ,\quad \omega _{o}^{\circ }=\sum_{n}\eta _{n}^{\circ
}\eta _{n}^{\circ \dagger }\otimes \eta _{n}^{\circ }\eta _{n}^{\circ
\dagger }\lambda _{0}\left( n\right) ,  \label{3.3}
\end{equation}
coinciding with (\ref{3.1}) in the case of Abelian $\mathcal{A}^{\circ }$.
These input-output compound states $\varpi $ are achieved by compositions $%
\lambda =\pi ^{\circ }\Lambda $, describing the entanglements $\lambda ^{*}$
of the extreme probe system $\left( \mathcal{B}^{\circ },\varsigma
_{0}\right) =\left( \mathcal{A}^{\circ },\varrho _{0}\right) $ to the output 
$\left( \mathcal{A},\varrho \right) $ of the channel.

If $\mathrm{K}:\mathcal{B}\rightarrow \mathcal{B}^{\circ }$ is a normal
completely positive unital map 
\begin{equation*}
\mathrm{K}\left( B\right) =\mathrm{tr}_{\mathcal{F}_{-}}X^{\dagger }BX,\quad
B\in \mathcal{B},
\end{equation*}
where $X$ is a bounded operator $\mathcal{F}_{-}\otimes \mathcal{G}
_0\rightarrow \mathcal{G}$ with $\mathrm{tr}_{\mathcal{F}_{-}}X^{\dagger
}X=I^{\circ }$, the compositions $\varkappa =\pi _{*}^{\circ }\mathrm{K}$, $%
\pi _{*}=\Lambda _{*}\varkappa $ are the entanglements of the probe system $%
\left( \mathcal{B},\varsigma \right) $ to the channel input $\left( \mathcal{%
\ A}^{\circ },\varrho _0\right) $ and to the output $\left( \mathcal{A}
,\varrho \right) $ via this channel. The state $\varsigma =\varsigma _0%
\mathrm{K}$ is given by 
\begin{equation*}
\mathrm{K}_{*}\left( \sigma _0\right) =X\left( I^{-}\otimes \sigma _0\right)
X^{\dagger }\in \mathcal{B}_{*}
\end{equation*}
for each density operator $\sigma _0\in \mathcal{B}_{*}^{\circ }$, where $%
I^{-}$ is the identity operator in $\mathcal{F}_{-}$. The resulting
entanglement $\pi _{*}=\lambda _{*}\mathrm{K}$ defines the compound state $%
\varpi =\varpi _0\left( \mathrm{K}\otimes \Lambda \right) $ on $\mathcal{B}
\otimes \mathcal{A}$ with 
\begin{equation*}
\varpi _0\left( B^{\circ }\otimes A^{\circ }\right) =\mathrm{tr}\tilde{B}
^{\circ }\pi ^{\circ }\left( A^{\circ }\right) =\mathrm{tr}\upsilon
_0^{\dagger }\left( B^{\circ }\otimes A^{\circ }\right) \upsilon _0.
\end{equation*}
on $\mathcal{B}^{\circ }\otimes \mathcal{A}^{\circ }$. Here $\upsilon _0:%
\mathcal{F}_0\rightarrow \mathcal{G}_0\otimes \mathrm{\mathcal{H}}_0$ is the
amplitude operator, uniquely defined by the input compound state $\varpi
_0\in \mathcal{B}_{*}^{\circ }\otimes \mathcal{A}_{*}^{\circ }$ up to a
unitary operator $U^{\circ }$ on $\mathcal{F}_0$, and the effect of the
input entanglement $\varkappa $ and the output channel $\Lambda $ can be
written in terms of the amplitude operator of the state $\varpi $ as 
\begin{equation*}
\upsilon =\left( X\otimes Y\right) \left( I^{-}\otimes \upsilon _0\otimes
I^{+}\right) U
\end{equation*}
up to a unitary operator $U$ in $\mathcal{F}=\mathcal{F}_{-}\otimes \mathcal{%
\ F}_0\otimes \mathcal{F}_{+}$. Thus the density operator $\omega =\upsilon
\upsilon ^{\dagger }$ of the input-output compound state $\varpi $ is given
by $\varpi _0\left( \mathrm{K}\otimes \Lambda \right) $ with the density 
\begin{equation}
\left( \mathrm{K}\otimes \Lambda \right) _{*}\left( \omega _0\right) =\left(
X\otimes Y\right) \omega _0\left( X\otimes Y\right) ^{\dagger },  \label{3.4}
\end{equation}
where $\omega _0=\upsilon _0\upsilon _0^{\dagger }$.

Let $\mathcal{K}_q$ be the convex set of normal completely positive maps $%
\varkappa :\mathcal{B}\rightarrow \mathcal{A}_{*}^{\circ }$ normalized as $%
\mathrm{tr}\varkappa \left( I\right) =1$, and $\mathcal{K}_q^{\circ }$ be
the convex subset $\left\{ \varkappa \in \mathcal{K}_q:\varkappa \left(
I\right) =\varrho _0\right\} $. Each $\varkappa \in \mathcal{K}_q^{\circ }$
can be decomposed as $\pi _{*}^{\circ }\mathrm{K}$, where $\pi _{*}^{\circ
}=\pi ^{\circ }$ is the standard entanglement on $\left( \mathcal{A}^{\circ
},\varrho _0\right) $, and $\mathrm{K}$ is a normal unital CP map $\mathcal{B%
}\rightarrow \mathcal{A}^{\circ }$. Further let $\mathcal{K}_c$ be the
convex set of the maps $\varkappa $, dual to the input maps of the form (\ref
{2.1}), described by the combinations 
\begin{equation}
\varkappa \left( B\right) =\sum_n\varsigma \left( B\right) \rho _0\left(
n\right) .  \label{3.6}
\end{equation}
of the primitive maps $\varkappa _n:B\mapsto \varsigma _n\left( B\right)
\rho _0\left( n\right) $, and $\mathcal{K}_d$ be the subset of the diagonal
decompositions 
\begin{equation}
\varkappa \left( B\right) =\sum_n\langle n|B|n\rangle \rho _0\left( n\right)
.  \label{3.5}
\end{equation}
As in the first case $\mathcal{K}_c^{\circ }$ and $\mathcal{K}_d^{\circ }$
denote the convex subsets corresponding to a fixed $\varkappa \left(
I\right) =\varrho _0$, and each $\varkappa \in \mathcal{K}_c^{\circ }$ can
be represented as $\pi _{*}^{\circ }\mathrm{K}$, where $\pi _{*}^{\circ }$
is a d-entanglement, which can be always be made pure by a proper choice of
the CP map $\mathrm{K}:\mathcal{B}\rightarrow \mathcal{A}^{\circ }$.
Furthermore let $\mathcal{K}_o$ ($\mathcal{K}_o^{\circ }$) be the subset of
all decompositions (\ref{3.6}) with orthogonal $\rho _0\left( n\right) $
(and fixed $\sum_n\rho _0\left( n\right) =\rho _0$): 
\begin{equation*}
\quad \rho _0\left( m\right) \rho _0\left( n\right) =0,\,m\neq n.
\end{equation*}
Each $\varkappa \in \mathcal{K}_o^{\circ }$ can be also represented as $\pi
_{*}^{\circ }\mathrm{K}$, where $\pi _{*}^{\circ }$ is a diagonal pure
o-entanglement $\mathcal{B}\rightarrow \mathcal{A}^{\circ }$.

Now, let us maximize the entangled mutual entropy for a given quantum
channel $\Lambda $ and a fixed input state $\varrho _{0}$ by means of the
above four types of compound states. The mutual entropy (\ref{4.3}) was
defined in the previous section by the density operators of the compound
state $\varpi $ on $\mathcal{B}\otimes \mathcal{A}$, and the product-state $%
\varphi =\varsigma \otimes \varrho $ of the marginals $\varsigma ,\varrho $
for $\varpi $. In each case 
\begin{equation*}
\varpi =\varpi _{0}\left( \mathrm{K}\otimes \Lambda \right) ,\quad \varphi
=\varphi _{0}\left( \mathrm{K}\otimes \Lambda \right) ,
\end{equation*}
where $\mathrm{K}$ is a CP map $\mathcal{B}\rightarrow \mathcal{B}^{\circ }$
, $\varpi _{0}$ is one of the corresponding extreme compound states $\varpi
_{q}^{\circ }$, $\varpi _{c}^{\circ }=\varpi _{d}^{\circ }$, $\varpi
_{o}^{\circ }$ on $\mathcal{A}^{\circ }\otimes \mathcal{A}^{\circ }$, and $%
\varphi _{0}=\varrho _{0}\otimes \varrho _{0}$. The density operator $\omega
=\left( \mathrm{K}\otimes \Lambda \right) _{*}\left( \omega _{0}\right) $ is
written in (\ref{3.4}), and $\phi =\sigma \otimes \rho $ can be written as 
\begin{equation*}
\phi =\varkappa _{*}\left( I\right) \otimes \lambda _{*}\left( I\right) ,
\end{equation*}
where $\lambda _{*}=\Lambda _{*}\pi _{*}^{\circ }$.

\bigskip

\noindent
{\bf Proposition 5.1.} 
{\it
The entangled mutual entropies achieve the following maximal values 
\begin{equation}
\sup_{\varkappa \in \mathcal{K}_{q}^{\circ }}I_{\mathcal{A},\mathcal{B}}\left( \varpi \right) =I_{q}\left( \varrho _{0},\Lambda \right) :=I_{\mathcal{A},\mathcal{A}^{\circ }}\left( \varpi _{q}^{\circ }\Lambda \right) ,
\label{3.7}
\end{equation}
\begin{equation*}
I_{c}\left( \varrho _{0},\Lambda \right) =\sup_{\varkappa \in \mathcal{K}_{c}^{\circ }}I_{\mathcal{A},\mathcal{B}}\left( \varpi \right) =\sup_{\varpi
_{d}^{\circ }}I_{\mathcal{A},\mathcal{A}^{\circ }}\left( \varpi _{d}^{\circ
}\Lambda \right) =I_{d}\left( \varrho _{0},\Lambda \right) ,
\end{equation*}
\begin{equation}
\sup_{\varkappa \in \mathcal{K}_{o}^{\circ }}I_{\mathcal{A},\mathcal{B}}\left( \varpi \right) =I_{o}\left( \varrho _{0},\Lambda \right)
:=\sup_{\varpi _{o}^{\circ }}I_{\mathcal{A},\mathcal{A}^{\circ }}\left(
\varpi _{o}^{\circ }\Lambda \right) ,  \label{3.8}
\end{equation}
where $\varpi _{\bullet }^{\circ }$ are the corresponding extremal input
entangled states on $\mathcal{A}^{\circ }\otimes \mathcal{A}^{\circ }$ with
marginals $\varrho _{0}$. They are ordered as 
\begin{equation}
I_{q}\left( \varrho _{0},\Lambda \right) \geq I_{c}\left( \varrho
_{0},\Lambda \right) =I_{d}\left( \varrho _{0},\Lambda \right) \geq
I_{o}\left( \varrho _{0},\Lambda \right) .  \label{3.9}
\end{equation}
}

\bigskip

We shall denote the maximal informations $I_c\left( \varrho _0,\Lambda
\right) =I_d\left( \varrho _0,\Lambda \right) $ simply as $I\left( \varrho
_0,\Lambda \right) $.

\bigskip

\noindent
{\bf Definition 5.1.} 
{\it
The supremums 
\begin{equation*}
C_{q}\left( \Lambda \right) =\sup_{\varkappa \in \mathcal{K}_{q}}I_{\mathcal{%
A},\mathcal{B}}\left( \varpi \right) =\sup_{\varrho _{0}}I_{q}\left( \varrho
_{0},\Lambda \right) ,\;
\end{equation*}
\begin{equation}
\sup_{\varkappa \in \mathcal{K}_{c}}I_{\mathcal{A},\mathcal{B}}\left( \varpi
\right) =C\left( \Lambda \right) :=\sup_{\varrho _{0}}I\left( \varrho
_{0},\Lambda \right) ,\;  \label{3.10}
\end{equation}
\begin{equation*}
C_{o}\left( \Lambda \right) =\sup_{\varkappa \in \mathcal{K}_{o}}I_{\mathcal{%
A},\mathcal{B}}\left( \varpi \right) =\sup_{\varrho _{0}}I_{o}\left( \varrho
_{0},\Lambda \right) ,\;
\end{equation*}
are called the q-, c- or d-, and o-capacities respectively for the quantum
channel defined by a normal unital CP map $\Lambda :\mathcal{A}\rightarrow 
\mathcal{A}^{\circ }$.
}

\bigskip

Obviously the capacities (\ref{3.10}) satisfy the inequalities 
\begin{equation*}
C_o\left( \Lambda \right) \leq C\left( \Lambda \right) \leq C_q\left(
\Lambda \right) .
\end{equation*}

\bigskip

\noindent
{\bf Theorem 5.2.} 
{\it
Let $\Lambda \left( A\right) =Y^{\dagger }AY$ be a unital CP map $\mathcal{A}%
\rightarrow \mathcal{A}^{\circ }$ describing a quantum deterministic
channel. Then 
\begin{equation*}
I\left( \varrho _{0},\Lambda \right) =I_{o}\left( \varrho _{0},\Lambda
\right) =S\left( \varrho _{0}\right) ,\quad I_{q}\left( \varrho _{0},\Lambda
\right) =\widetilde{S}\left( \varrho _{0}\right) ,
\end{equation*}
and thus in this case 
\begin{equation*}
C\left( \Lambda \right) =C_{o}\left( \Lambda \right) =\ln \mathrm{rank}%
\mathcal{A}^{\circ },\quad C_{q}\left( \Lambda \right) =\ln \dim \mathcal{A}%
^{\circ }
\end{equation*}
}

\bigskip

In the general case d-entanglements can be more informative than
o-entanglements as it can be shown on an example of a quantum noisy channel
for which 
\begin{equation*}
I\left( \varrho _{0},\Lambda \right) >I_{o}\left( \varrho _{0},\Lambda
\right) ,\quad C\left( \Lambda \right) >C_{o}\left( \Lambda \right) .
\end{equation*}
The last equalities of the above theorem will be related to the work on
entropy by Voiculescu \cite{Voi}.

\end{document}